*Article*

# A MEMS-based terahertz broadband beam steering technique

Weihua Yu [1,2]*, Hong Peng [2], Mingze Li[1,2], Haolin Li[1], Yuan Xue [2], Huikai Xie [1,2]

1. School of Integrated Circuits and Electronics, Beijing Institute of Technology, Beijing 100081, China; ywhbit@bit.edu.cn (W.Y.); limingze@bit.edu.cn (H.L.); lihaolin@bit.edu.cn (H.L.); hk.xie@ieee.org (H.X.);.
2. BIT Chongqing Institute of Microelectronics and Microsystems, Chongqing 401332, China; xueyuan@ccmems.cn (Y.X.)
* Correspondence: ywhbit@bit.edu.cn (W.Y.)

**Abstract:** A multi-level tunable reflection array wide-angle beam scanning method is proposed to address the limited bandwidth and small scanning angle issues of current terahertz beam scanning technology. In this method, a focusing lens and its array are used to achieve terahertz wave spatial beam control, and MEMS mirrors and their arrays are used to achieve wide-angle beam scanning. The 1~3 order terahertz MEMS beam scanning system designed based on this method can extend the mechanical scanning angle of MEMS mirrors by 2~6 times, when tested and verified using an electromagnetic MEMS mirror with a 7mm optical aperture and a scanning angle of 15° and a D-band terahertz signal source. The experiment shows that the operating bandwidth of the first-order terahertz MEMS beam scanning system is better than 40GHz, the continuous beam scanning angle is about 30°, the continuous beam scanning cycle response time is about 1.1ms, and the antenna gain is better than 15dBi at 160GHz. This method has been validated for its large bandwidth and scalable scanning angle, and has potential application prospects in terahertz dynamic communication, detection radar, scanning imaging, and other fields.

**Keywords:** terahertz; beam steering; MEMS mirror; antenna ; dielectric lens.

## 1. Introduction

Beam scanning technology can accurately control and locate the direction of electromagnetic wave beams, and is one of the core technologies used in various detection radars, dynamic communication, and scanning imaging systems[1]. In the microwave and millimeter wave frequency bands, beam scanning technology is mainly achieved through mechanical scanning antennas, phased arrays, and multi beam antennas[2], while in the infrared and visible light frequency bands, beam scanning technology is usually achieved through mechanical scanning mirrors, electro-optic scanning, acousto-optic scanning, and other technologies[3-4].

Currently, with the rapid development of terahertz technology in fields such as 6G communication, integrated sensing, biomedical imaging, and security control, higher communication rates and detection resolutions have put forward higher requirements for the working bandwidth of terahertz beam scanning. At present, low-cost passive frequency beam scanning technology is no longer applicable[5-10]. Therefore, researchers have conducted extensive research on broadband terahertz beam scanning technology based on traditional phased array technology based on semiconductor devices [11-13] and quasi optical scanning technology based on mechanical scanning [14-16]. However, due to the low operating frequency band, poor PIN switching performance, and high losses of terahertz phase shifters, the implementation cost of terahertz phased arrays is high and the scanning angle is small; The beam scanning scheme based on mechanical structure, although having good single beam performance and large scanning angle, lacks the problems of large volume and slow scanning speed.

In addition, some studies have also explored terahertz transmission and reflection metasurface beam scanning methods based on new materials such as liquid crystals [17-18] and graphene [19-20]. However, this technology not only has complex design, high cost, but also small beam scanning angles and limited bandwidth.

In order to fully utilize the advantages of terahertz bandwidth and support the development of higher communication rates and resolutions in terahertz communication and radar systems, it is urgent to explore a terahertz beam scanning implementation scheme with large working bandwidth and scanning angle.

In this article, a multi-level tuning reflection array wide-angle beam scanning method is first proposed. Through the analysis of the spatial beam manipulation ability of the dielectric lens and metal reflector, a terahertz 1~3 order terahertz MEMS beam scanning system is designed based on the dielectric lens and MEMS reflector, which can achieve beam scanning of ±15°, ±30°, and ±45°, respectively; Then, by verifying the processing of dielectric lenses and the selection of MEMS reflectors, a first-order terahertz MEMS beam scanning system and its testing environment were built, and the terahertz MEMS beam scanning system was constructed and performance tests such as working bandwidth, scanning angle, and beam gain were completed. The validation of the new high bandwidth and high angle beam scanning method has been achieved.

## 2. Terahertz Multi stage Tuned Reflective Array Beam Scanning Technology

A multi-level reflector beam scanning method based on quasi optical theory was studied to address the current issues of small scanning angles and narrow working bandwidth in terahertz waves. At the same time, the spatial beam control capabilities of the medium lens and reflector were utilized to solve the problems of beam diffusion and low transmission efficiency in this method.

*2.1. The principle of beam scanning angle extension for multi-level tuning reflection array*

As shown in Figure 1(a), the schematic diagram of a single reflection beam scanning shows that when the mechanical scanning angle of the reflector is $\pm A0$, the beam emitted from the laser will deflect through the reflector, resulting in a scanning angle of $\pm A1$. According to the law of reflection, it can be concluded that A3=180-2*A2; A3+A1=180-2*(A2-A0); Therefore, A1=2*A0.

The schematic diagram of the secondary reflection beam scanning shown in Figure 1 (b) shows that when a single reflection beam passes through three mirrors located in the same plane and perpendicular to the center beam for secondary reflection, if the mechanical scanning angles of all three mirrors are $\pm A0$, the beam scanning angle after secondary reflection can be calculated as $\pm(90°-B1+A1)$; According to the reflection law, B1=90°-A1, therefore the scanning range of the secondary reflection beam is $\pm 2*A1=\pm 4*A0$;

The schematic diagram of the secondary reflection beam scanning shown in Figure 1 (c) shows that when the secondary reflection beam is reflected by 7 mirrors located in the same plane and perpendicular to the secondary reflection center beam, if the mechanical scanning angle of the 7 mirrors is also $\pm A0$, the beam scanning angle after three reflections can be calculated as $\pm(90°-C1+A1)$; According to the reflection law, C1=B1-A1, therefore the scanning range of the secondary reflection beam is $\pm(90°-B1+2*A1)=\pm 3*A1=\pm 6*A0$;

Similarly, for N reflections, if the mechanical scanning angle of the reflecting mirror for each reflection is $\pm A0$, and the array of reflecting mirrors for each reflection is in the same plane and perpendicular to the center beam of N-1 reflections, then the scanning angle of the beam after the Nth reflection is $\pm(90° - B1+(N-1) A1)=\pm N*A1=\pm 2N*A0$; Due to B1 and A1 being positive, and the theoretical maximum scanning angle of this method being $\pm 90°$.

Based on the above analysis, if the mechanical scanning angle of the reflector is ±7.5°, then the scanning angle for one reflection is ±15°; The scanning angle for two reflections is ±30°; The scanning angle for three reflections is ±45°;

Of course, when using this method for terahertz waves, due to the wide beam of terahertz waves emitted by the antenna feed source, there is inevitably a problem of terahertz beam divergence, which will lead to low beam transmission efficiency during multi-level reflection; To solve this problem, on the one hand, it is necessary to enhance the gain of terahertz beams, reduce the divergence angle, and at the same time, design a terahertz beam focusing lens to achieve efficient transmission of terahertz beams during multi-level reflection.

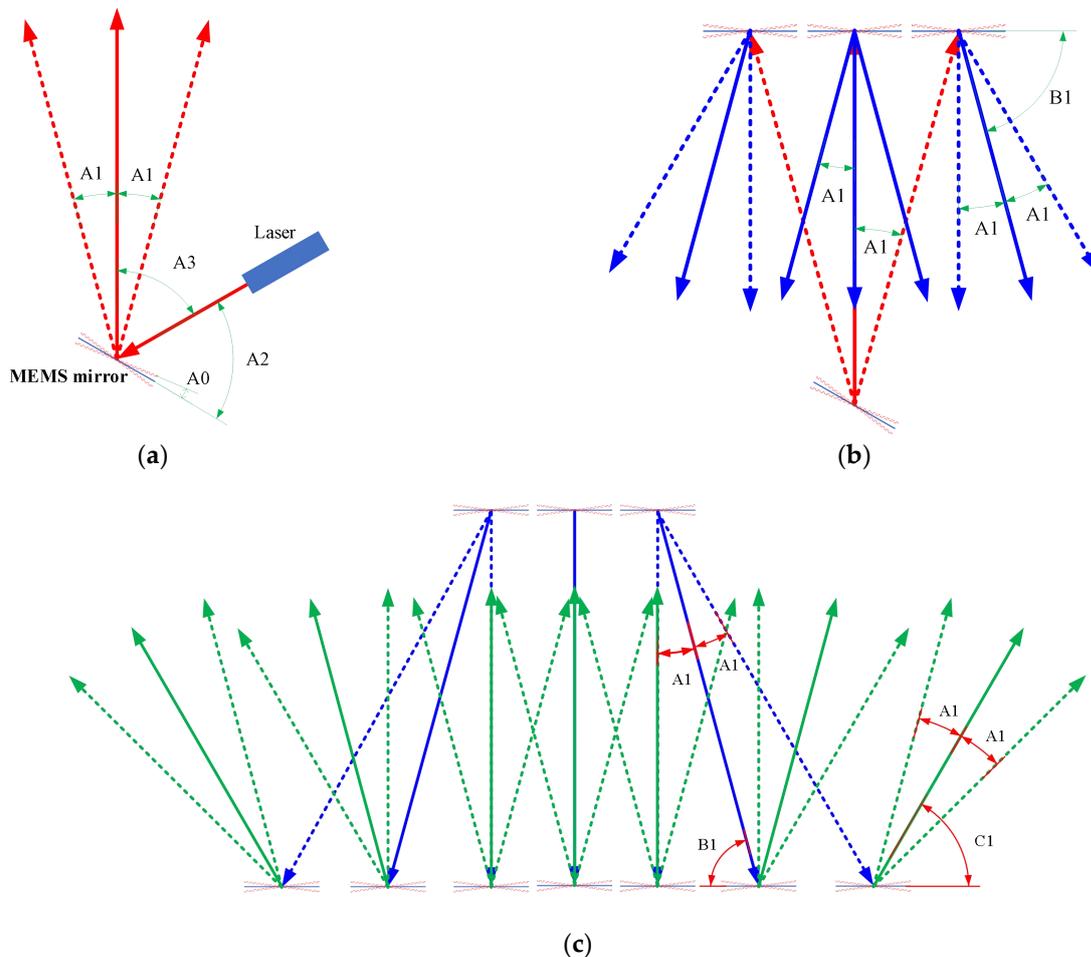

Figure 1. MEMS-based THz beam steering: (a) Single reflection beam scanning; (b) Secondary reflection beam scanning; (c) Triple reflection beam scanning.

*2.2. The Influence of Terahertz Lens on Beam*

In order to achieve efficient focusing of terahertz waves, design of terahertz dielectric lenses has been carried out. Due to the types of dielectric lenses such as spherical lenses, hyperbolic lenses, Ryomb lenses, and free-form surface shaping lenses, a simple and efficient design method for spherical lenses is introduced below.

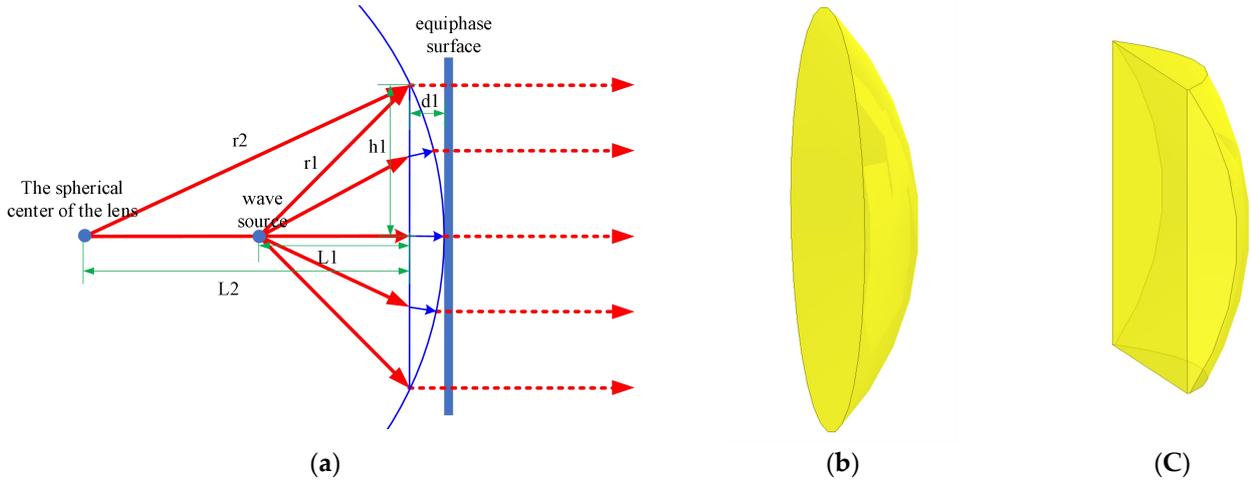

(a) (b) (C)

**Figure 2.** Spherical collimating lens: (a) Design principle diagram; (b) Circular lens; (c) Square lens.

As shown in Figure 1 (a), if the relative dielectric constant of the dielectric lens is $\varepsilon_r$, the mirror thickness d1 of a spherical collimating lens with a focal length of L1 and a radius of h1 satisfies the following formula:

$$d1 = \frac{r1 - L1}{\sqrt{\varepsilon_r} - 1} = \frac{\sqrt{L1^2 + h1^2} - L1}{\sqrt{\varepsilon_r} - 1} \quad (1)$$

Meanwhile, the reference circle radius r2 of the spherical collimating lens satisfies the formula:

$$r2 = \frac{d1^2 + h1^2}{2 * d1} \quad (2)$$

Based on the above two formulas, a spherical collimating lens with a focal length of L1 and a radius of h1 can be quickly achieved, as shown in Figure 1 (b). When designing a lens using a photosensitive resin material with a dielectric constant of $\varepsilon_r$ =2.739, a focal length of 7mm can be achieved; A spherical collimating lens with a radius of 4mm, with a mirror thickness d1 ≈ 2.273mm; The radius of the reference circle r2≈7.2mm.

To verify the performance of the spherical lens, a rectangular waveguide antenna was designed as shown in Figure 3 (a), and the simulation results at 160GHz are shown in Figure (b): its gain is 7.9dBi; The 3dB beam width is ±30°.

When the above-mentioned dielectric lens is loaded in front of the rectangular waveguide, a dielectric lens antenna as shown in Figure (c) can be formed, where the size of the dielectric lens is 8 × 8mm; Focal length 7mm; 7mm from the rectangular waveguide interface; The antenna can receive a beam energy of ±30° from a rectangular waveguide. The simulation results at 160GHz are shown in Figure (d), with a gain of 20.1dBi; The 3dB beam width is ±5.1°; Visible medium lenses have good focusing ability for terahertz beams and can significantly enhance beam gain; This is very important for the multi-level reflection beam scanning method proposed above.

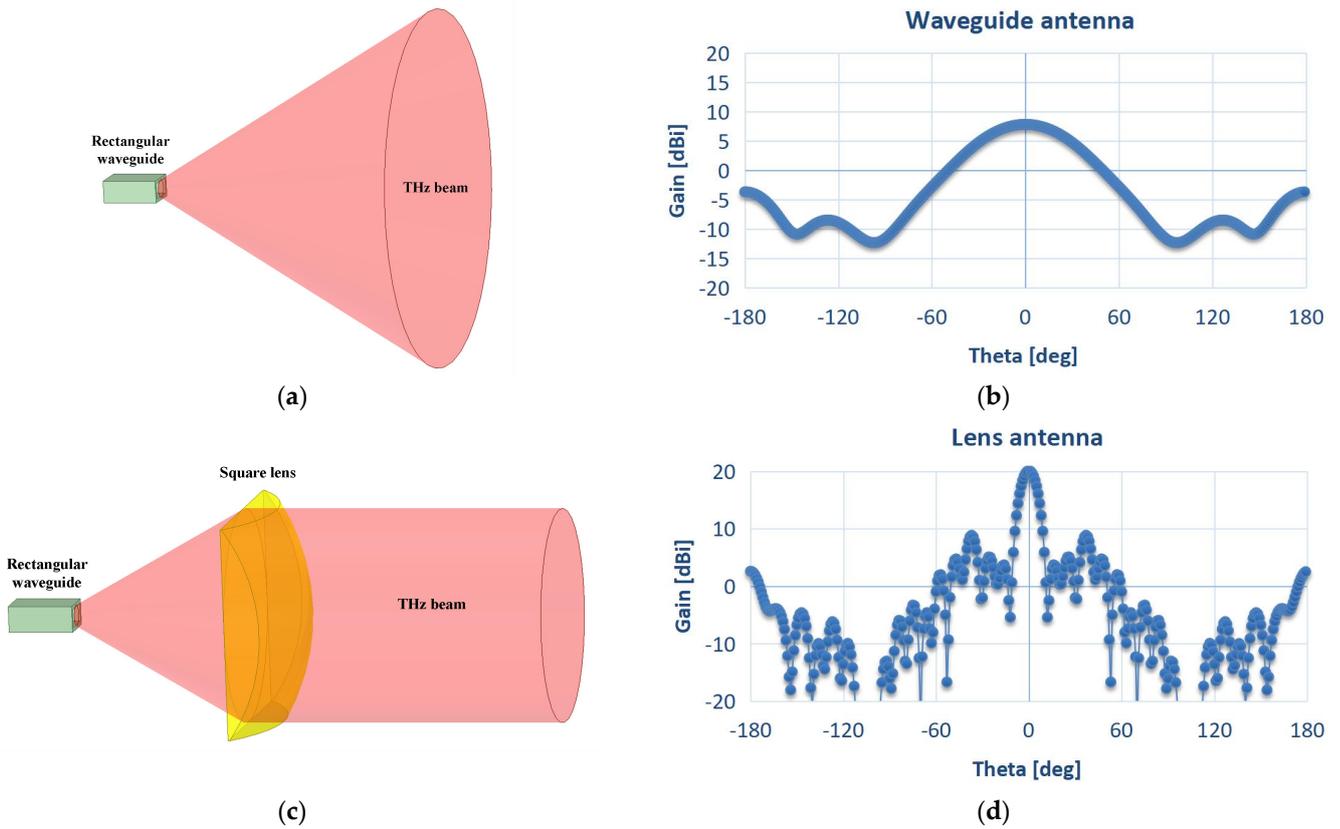

**Figure 3.** Lens antenna based on spherical collimating lens and its simulation results: (a) Structural diagram of waveguide antenna; (b) Gain curve of 160GHz waveguide antenna at phi=0; (c) Structural diagram of lens antenna; (d) Gain curve of 160GHz lens antenna at phi=0; .

*2.3. The influence of mirror size on the beam*

In the terahertz frequency band, due to the influence of aperture on terahertz antennas, loading a metal reflector in front of the terahertz antenna will greatly affect antenna gain and beam width. As shown in Figure 4 (a), a metal reflector with a height of L and a width of 1.2*L is placed at the front end of the lens antenna, where the reflector is 10mm away from the lens; The angle between the reflector and the horizontal terahertz beam is 30°, and the angle between the terahertz beam reflected by the reflector and the horizontal terahertz beam is 60°;

When the dimensions of the mirrors are 6×7.2mm respectively; 7×8.4mm; 8×9.6mm; 10×12mm; At a size of 12×14.4mm, the simulation results are shown in Figure 4 (b), where different mirrors achieve maximum gain at theta=60deg and maximum sidelobe level at 180deg.

Among them, when the reflector is 6×7.2mm, the maximum gain of the reflected beam is 13.9dBi, the 3dB beam width is 10.3deg, and the sidelobe level is -1.4dB (72.44%); When the reflector is 7×8.4mm, the maximum gain of the reflected beam is 15.4dBi, the 3dB beam width is 10.4deg, and the sidelobe level is -5.1dB (30.9%); When the reflector is 8×9.6mm, the maximum gain of the reflected beam is 16.3dBi, the 3dB beam width is 10.2deg, and the sidelobe level is -8.5dB (14.13%); When the reflector is 10×12mm, the maximum gain of the reflected beam is 17.7dBi, the 3dB beam width is 9.4deg, and the sidelobe level is -11.6dB (6.9%); When the reflector is 12×14.4mm, the maximum gain of the reflected beam is 18.5dBi, the 3dB beam width is 9.1deg, and the sidelobe level is -11.9dB (6.46%);

Based on the sidelobe level and maximum beam gain, the reflection efficiency of the four mirrors can be calculated as 58%, 76.4%, 87.62%, 93.54%, and 93.93%, respectively.

It can be observed that as the area of the reflector increases, the gain of the reflected beam increases and the reflection efficiency increases.

As shown in Figure 4 (c), in order to solve the problem of reflection efficiency, it is considered to change the original collimating lens to a focusing lens; When the reflector is 6×7.2mm, the simulation results of the medium lens focal lengths of 5.5mm, 6mm, 6.5mm, 7mm, 8mm, and 9mm are shown in Figure 4 (d).

When the focal length of the dielectric lens changes from 9mm to 5.5mm, the sidelobe gains in the Theta=180° direction are 13.9dBi, 13.62dBi, 12.47dBi, 10.7dBi, 7.76dBi, and 3.15dBi, respectively; The main lobe gains near Theta=60° are 12.6dBi, 13dBi, 13.9dBi, 13.2dBi, 12.15dBi, and 10.74dBi, respectively.

Meanwhile, compared to the main lobe near Theta=60°, the maximum sidelobe levels are+1.3dB,+0.6dB, -1.4dB, -2.5dB, -4.4dB, and -4.4dB, respectively; It can be roughly calculated that the beam reflection efficiency of the metal surface is 42.57%, 46.55%, 58%, 64%, 73.4%, and 73.4%, respectively. Therefore, when the size of the reflection surface is limited, greater reflection efficiency can be achieved by adjusting the focal length of the lens.

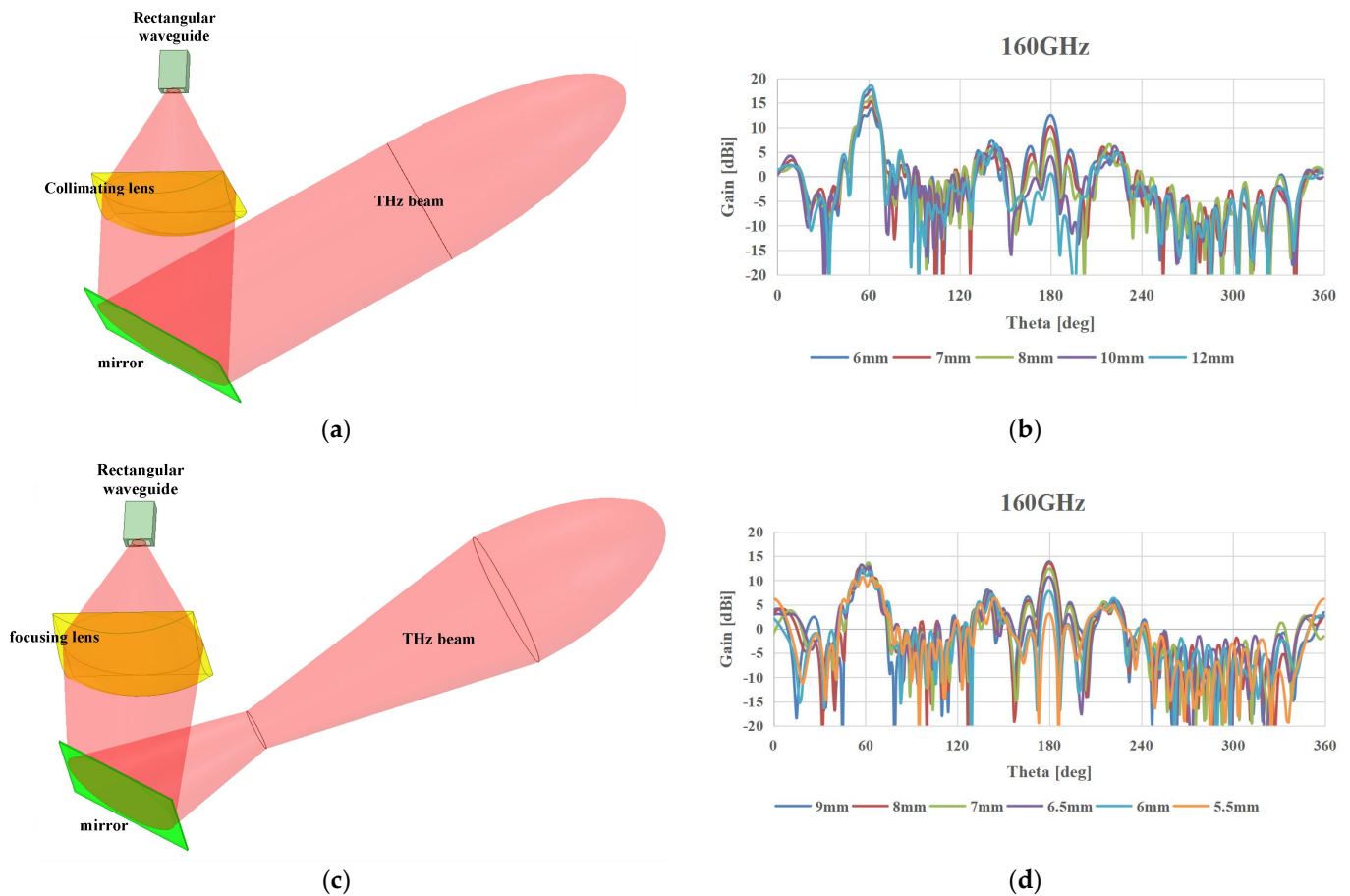

**Figure 4.** Analysis of terahertz beam reflection efficiency: (a) Principle of terahertz wave reflection based on collimating lenses; (b) The influence of different reflection surface sizes on the reflected beam at 160GHz; (c) The principle of terahertz wave reflection based on focusing lenses; (d) The effect of different lens focal lengths on reflected beams at 160GHz

## 3. Terahertz beam scanning technology based on MEMS mirrors

According to the analysis of the multi-level reflection theory mentioned above, by tuning the multi-level reflection of the reflector, a wide range of beam scanning angles can be achieved. At the same time, the problem of low efficiency and low gain in beam reflection can be solved through focusing lenses. Based on a 7mm MEMS reflector and an 8×8mm lens; Designed and simulated a 1~3 order beam scanning antenna system.

*3.1. Selection and testing of MEMS mirror*

In order to achieve terahertz wide angle beam scanning, according to the multi-level tunable reflection wide angle beam scanning method mentioned above, the larger the size of the reflector, the higher the transmission efficiency. The larger the mechanical scanning angle of the reflector, the larger the beam scanning angle. Therefore, a 7×8mm elliptical structure MEMS electromagnetic micro mirror chip was selected as the reflector, as shown in Figure 5 (A).

A testing environment for MEMS electromagnetic micro mirror chips was established using signal sources, DC power supplies, lasers, display boards, and other devices, as shown in (b) of Figure 5;

When 60Hz and 890Hz 1vPP sine signals are applied to the slow axis and fast axis of the MEMS chip respectively, the scanning pattern on the display board is obtained; The scanning speed of electromagnetic MEMS chips is better than 1.13ms;

When a DC signal of -3.5~3.5V is applied to the MEMS chip, as shown in the schematic diagram in Figure 5 (c), the laser emitted by the laser moves within the range of -25.1cm to 25.5cm through the reflection point of the MEMS chip, and the calculated laser deflection angle is -15.2~14.6°; By obtaining a deflection extreme value around ±3V, MEMS can achieve a mechanical scanning angle of approximately ±7.5° through a DC bias voltage of ±3V.

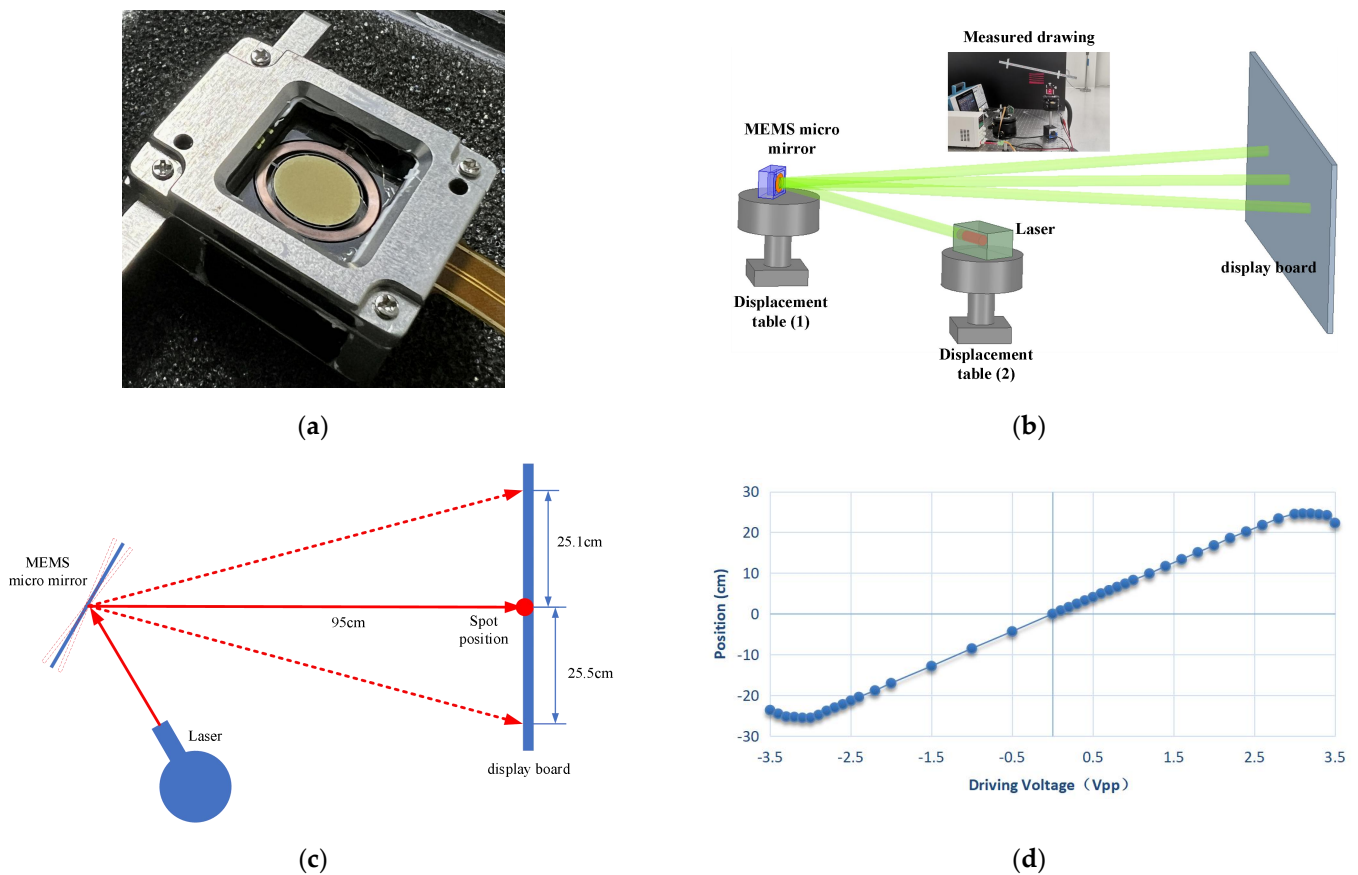

**Figure 5.** MEMS electromagnetic micro mirror chip testing (a) MEMS electromagnetic micro mirror chip structure; (b) MEMS micro mirror testing environment; (c) MEMS deflection angle testing plan; (d) Schematic diagram of the movement of reflected light spot on MEMS chips with DC bias voltage.

*3.2. Design and Simulation of 1~3 Order Terahertz MEMS Beam Scanning System*

As shown in Figure 6 (a), a first-order beam terahertz beam scanning system is designed based on a 7×8mm MEMS reflector and focusing lens. The focusing lens is a square structure with a size of 8×8mm, a focal length of 8.5mm, and a material of 2.739 photosensitive resin. The output signal of the terahertz source is in the 110-170GHz frequency band, and the output waveguide is a standard D-band waveguide port: WR-6.5 (1.65×0.83mm) standard rectangular waveguide. The MEMS reflector includes a metal mirror surface and driving components, placed at a 30° angle with the focusing lens, and the distance between the MEMS reflector and the focusing lens is 15mm; The focusing lens is 10mm away from the waveguide port.

When the metal mirror surface of the MEMS reflector deviates by ±15°, the simulation results of the first-order terahertz MEMS beam scanning based on CST are shown in Figure 6 (b-d).

At 120GHz, the scanning angle of the terahertz wave is 44°~75°, and the maximum gains of the five beams are 12.26dBi, 12.46dBi, 11.94dBi, 12.49dBi, 11.73dBi, and the average gain is about 12.18dBi;

At 140GHz, the scanning angle of Taihe wave is 46°~75°, and the maximum gains of the five beams are 13.29dBi, 14.64dBi, 14.49dBi, 14.2dBi, and 14.61dBi, respectively, with an average gain of about 14.25dBi;

At 160GHz, the scanning angle of Taihe wave is 43°~74°, and the maximum gains of the five beams are 15.88dBi, 15.88dBi, 16.31dBi, 16.04dBi, and 15.59dBi, respectively, with an average gain of about 15.94dBi.

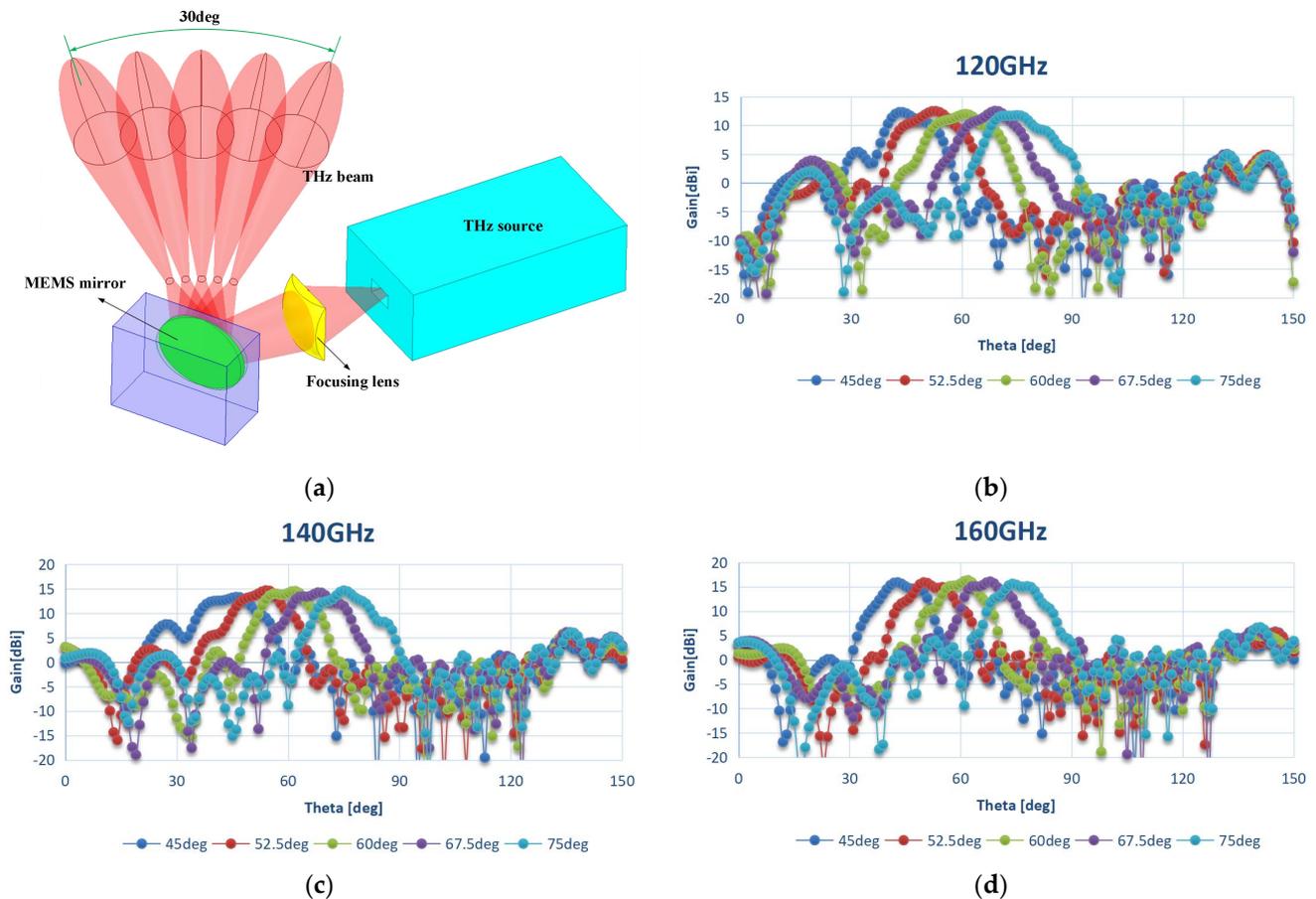

**Figure 6.** Design and Simulation of 1-Order Terahertz MEMS Beam Scanning System: (a) MEMS beam scanning antenna structure; (b) Simulation results at 120GHz; (c) Simulation results at 140GHz; (d) Simulation results at 160GHz.

As shown in Figure 7 (a), based on the multi-level tunable reflection array beam scanning method, a second-order terahertz beam scanning system was designed using a 1×3 lens array I and a 1×3 MEMS reflector array I on the basis of a first-order terahertz beam scanning system. The three focusing mirrors in lens array I are identical, with a size of 8×8mm, a focal length of 8.5mm, and a material of 2.739 photosensitive resin. The three lenses are perpendicular to the 45°, 60°, and 75° beam directions of the first-order terahertz beam scanning system. The MEMS mirrors in the 1×3 MEMS mirror array I are also identical, with a metal mirror size of 7×8mm ellipse; The mechanical deflection angle is ±7.5°, and the three MEMS reflectors are in the same plane perpendicular to the 60° beam direction of the first-order terahertz beam scanning system.

When the MEMS reflector of the first-order terahertz beam scanning system deflects discretely in three directions:+7.5°, 0°, and -7.5°, while the MEMS reflector of the 1×3 MEMS reflector array I deflects continuously in the range of -7.5° to 7.5°, it can achieve continuous scanning of terahertz waves within a certain range. The scanning results of the 2nd order terahertz MEMS beam based on CST are shown in Figure 7 (b-d),

At 120GHz, the scanning angle of Taihe wave is 29°~91°, and the average gain of 9 beams is about 11.37dBi;

At 140GHz, the scanning angle of Taihe wave is 31°~90°, and the average gain of 9 beams is about 13.46dBi;

At 160GHz, the scanning angle of Taihe wave is 29°~89°, and the average gain of 9 beams is about 15.14dBi.

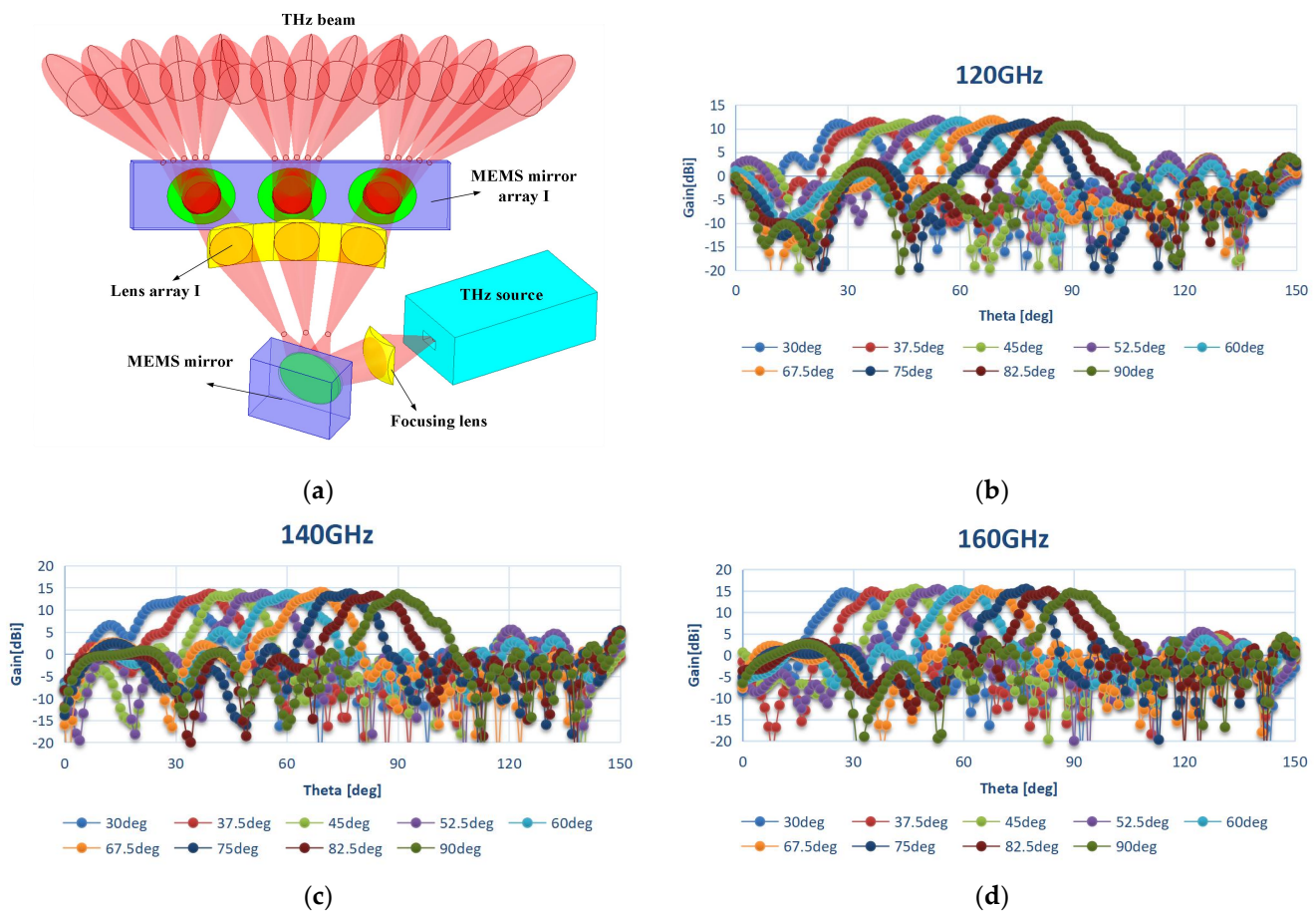

**Figure 7.** Design and Simulation of 2-Order Terahertz MEMS Beam Scanning System: (a) MEMS beam scanning antenna structure; (b) Simulation results at 120GHz; (c) Simulation results at 140GHz; (d) Simulation results at 160GHz.

As shown in Figures 8 (a) and 8 (b), based on the multi-level tunable reflection array beam scanning method, a third-order terahertz beam scanning system was designed using a 1×7 lens array II and a 1×7 MEMS reflector array II on the basis of a second-order terahertz beam scanning system. The seven focusing lenses in the lens array II are identical, with a size of 8×8mm and a focal length of 8.5mm. The material used is 2.739 photosensitive resin, and the seven lenses are perpendicular to the 30°, 45°, 60°, 60°, 75°, and 90° beam directions of the second-order terahertz beam scanning system. The MEMS mirrors in the 1×7 MEMS mirror array II are also identical, with a metal mirror size of 7×8mm ellipse; The mechanical deflection angle is ±7.5°, and the seven MEMS mirrors are in the same plane perpendicular to the 60° beam direction of the second-order terahertz beam scanning system.

When the MEMS reflector of the first-order terahertz beam scanning system deflects discretely in three directions:+7.5°, 0°, and -7.5°, the middle MEMS reflector of the second-order terahertz beam scanning system does not deflect, and the two end MEMS reflectors deflect discretely in three directions:+7.5°, 0°, and -7.5°; When the MEMS reflector of the 1×7 MEMS reflector array II is continuously deflected within the range of -7.5° to 7.5°, it can achieve continuous scanning of terahertz waves over a larger range. The scanning results of the third-order terahertz MEMS beam based on CST are shown in Figure 7 (c-e),

At 120GHz, the scanning angle of Taihe wave is 14°~105°, and the average gain of 9 beams is about 10.98dBi; At 140GHz, the scanning angle of Taihe wave is 16°~104°, and the average gain of 9 beams is about 13.07dBi; At 160GHz, the scanning angle of Taihe wave is 13°~104°, and the average gain of 9 beams is about 14.74dBi.

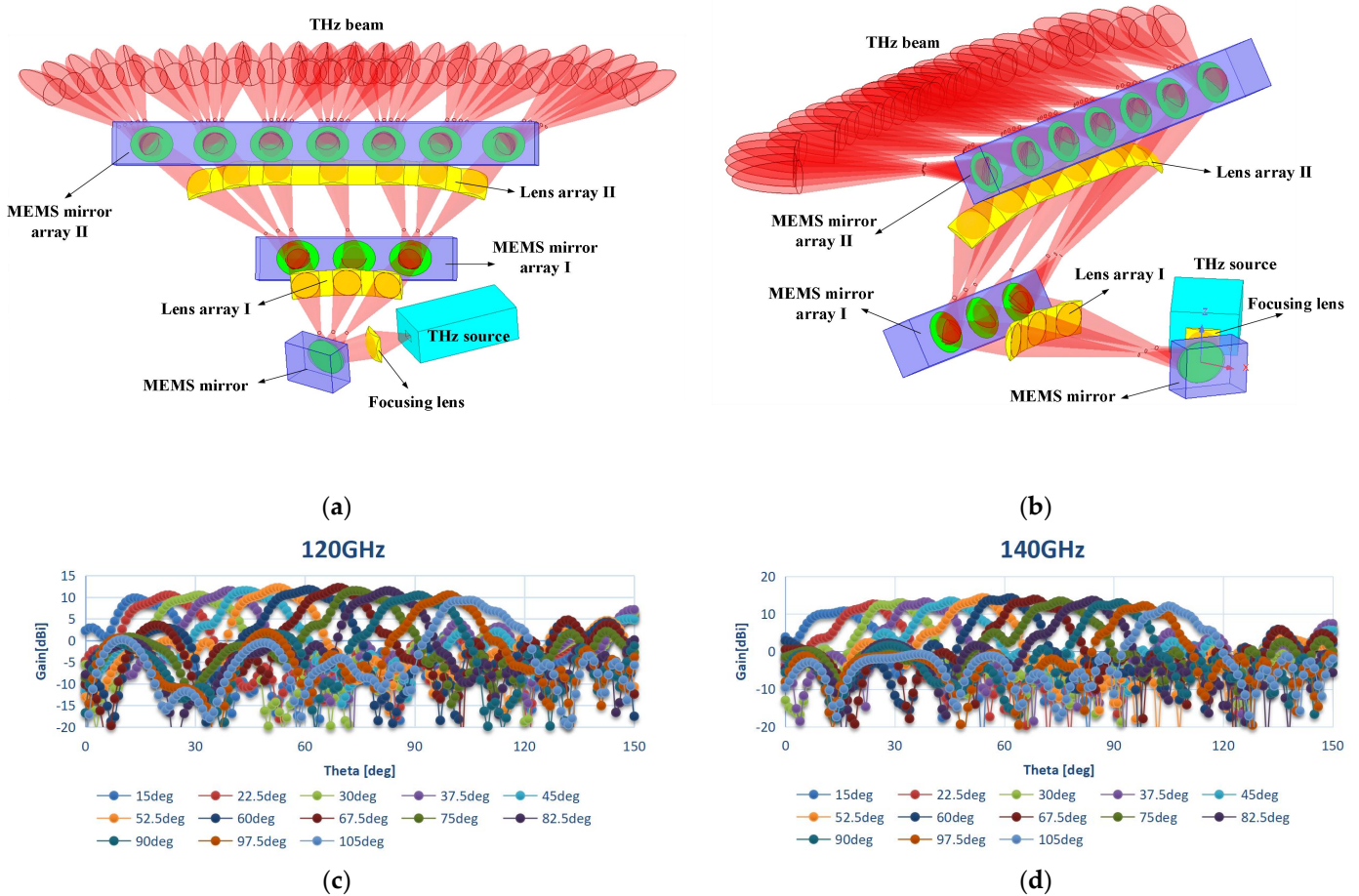

(a)　　　　　　　　　　　　　　　　　　　(b)

(c)　　　　　　　　　　　　　　　　　　　(d)

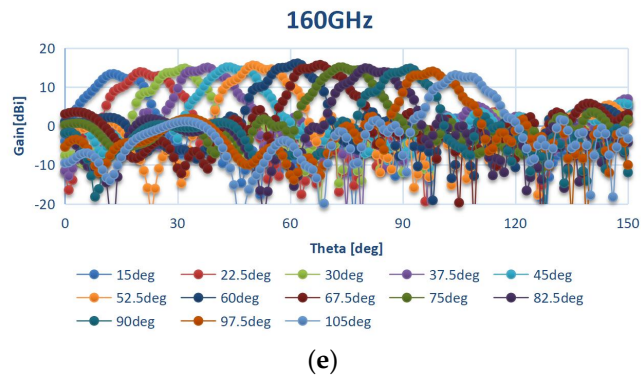

(**e**)

**Figure 8.** Design and Simulation of 3-Order Terahertz MEMS Beam Scanning System: (a) Front view of a third-order beam scanning antenna system; (b) Side view of a third-order beam scanning antenna system; (c) Simulation results at 120GHz; (d) Simulation results at 140GHz; (e) Simulation results at 160GHz.

## 4. Construction and verification of terahertz MEMS beam scanning testing system

In order to verify the beam scanning capability of the terahertz MEMS quasi optical beam scanning system, a first-order terahertz MEMS beam scanning testing system was constructed using D-band spectrophotometer, terahertz signal source, 7 mm MEMS electromagnetic micro mirror chip, 8mm focusing lens, receiving horn antenna, testing turntable, and other equipment as shown in Figure 9 (a).

Among them, the size of the focusing lens is 8×8mm; The focal length is 8.5mm, and the processing model is shown in Figure 9 (b). Using a photosensitive resin with a dielectric constant of 2.739, the physical image of the focusing lens processed using a 10um precision 3D printing process is shown in Figure 9 (c).

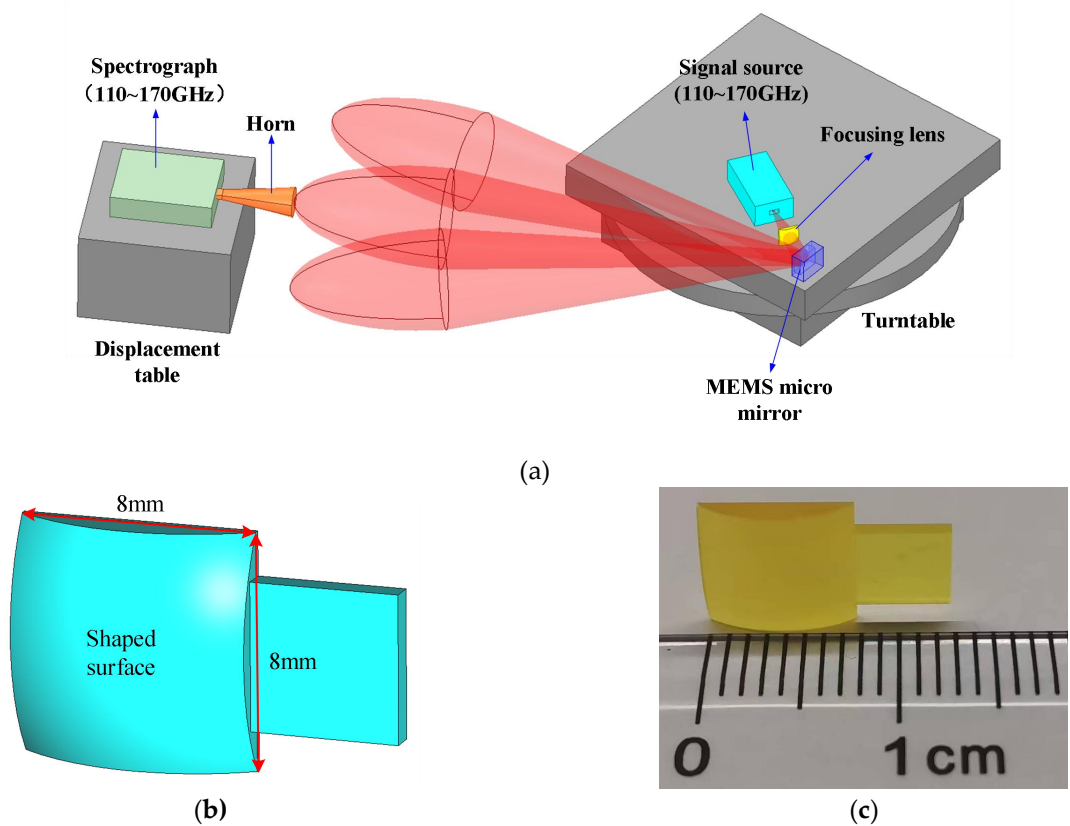

**Figure 9.** Test environment: (a) 1-Order Terahertz MEMS Beam Scanning Test System; (a) Optimize the machining model of focusing lens; (b) Physical image of focusing lens.

The physical image of the first-order terahertz MEMS beam scanning test system and the details of the focusing lens and MEMS reflector are shown in Figure 10 (a), where the waveguide port of the terahertz signal source is about 10mm away from the focusing lens; The distance between the waveguide port and the center of the MEMS electromagnetic micro mirror surface is approximately 25mm; By applying a -3V~+3V driving level to the MEMS reflector and continuously rotating the turntable, the directional pattern of the test terahertz wave is obtained.

Finally, the beam scanning results were obtained as shown in Figure 10 (b-d); When the frequency of the emitted terahertz wave is 120GHz, the maximum gain direction of the terahertz wave reflected by the MEMS reflector changes by 28°, achieving beam scanning in the range of 46~74°. The average gain is 11.56dBi, which is 0.62dB lower than the simulation value.

When the frequency of transmitting terahertz waves is 140GHz, the maximum gain direction of the terahertz beam reflected by MEMS changes by 30°, and beam scanning is achieved in the range of 45-75°. The average gain is 13.47dBi, which is 0.78dB lower than the simulation value.

When the frequency of transmitting terahertz waves is 160GHz, the maximum gain direction of the terahertz beam reflected by MEMS changes by 29°, and the beam scanning range ranges from 45° to 74°. The average gain is 15.12dBi, which is 0.82dB lower than the simulation value.

The beam scanning range and average gain of the above three frequency points are consistent with the first-order terahertz MEMS beam scanning results in Figure 6, verifying the correctness of the multi-level tunable reflection array wide-angle beam scanning method proposed in this paper.

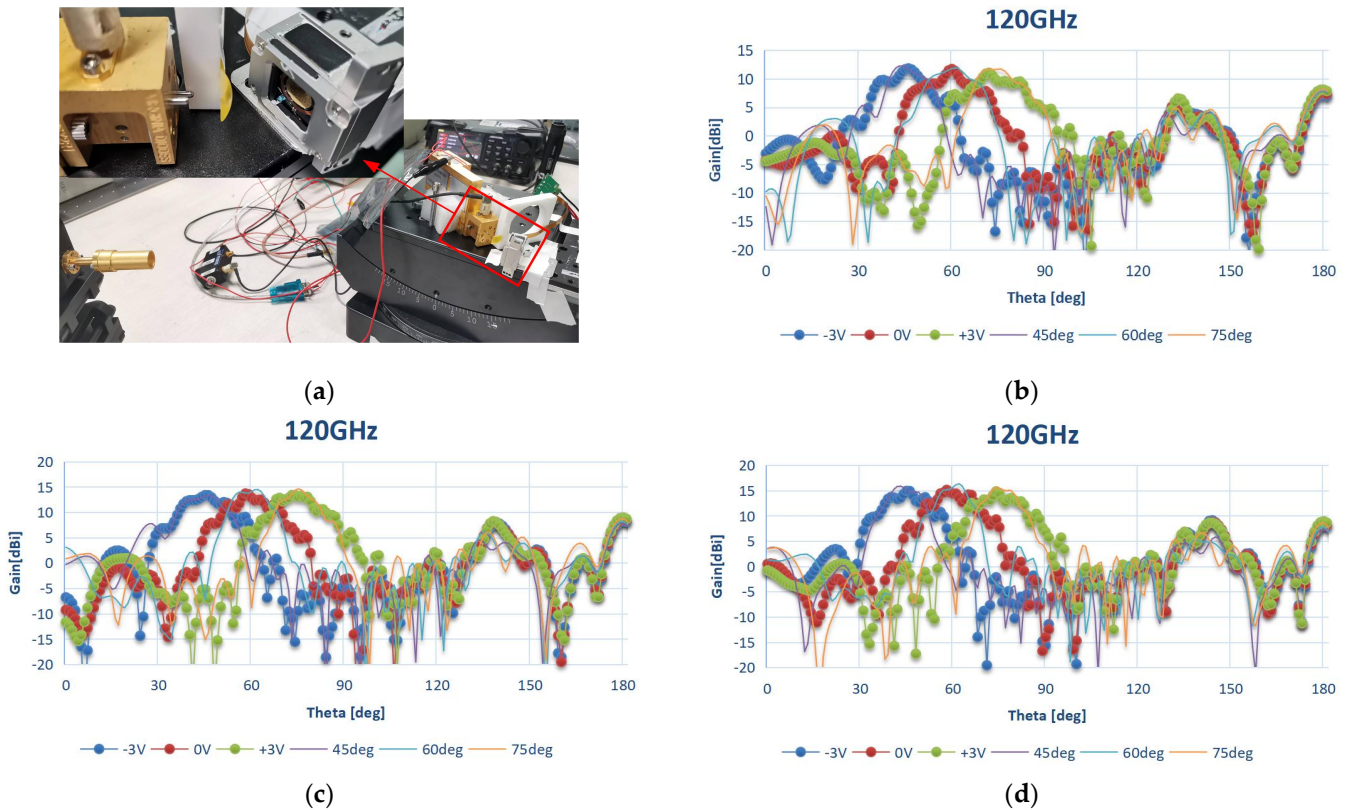

**Figure 10.** Single chip beam scanning test results: (a) Beam scanning antenna details; (b) Measured results at 120GHz; (c) Measured results at 140GHz; (d) Measured results at 160GHz.

In summary, the terahertz MEMS beam scanning antenna based on MEMS electromagnetic micro mirror chips and focusing lenses has a bandwidth of ≥ 40GHz in the D-band. Based on the existing 7mm MEMS electromagnetic micro mirror chips, the first level reflection can achieve a beam scanning range of 30°, the second level reflection can achieve a beam scanning range of 60°, the third level reflection can achieve a beam scanning range of 90°, and the scanning cycle speed is better than 1.13ms. The gain at 160GHz is better than 15dBi, and the scanning gain change is less than 1dB; Compared with current relevant literature, our work not only has the ability of continuous beam scanning, but also has a larger working bandwidth and scalable wide beam scanning range.

Table 1. Comparisons with Other Published Works

| Ref. | Design method | Center frequency | Bandwidth | scanning type | scanning speed | scanning range |
|---|---|---|---|---|---|---|
| [11] | Semiconductor | 320GHz | 6.25% | Continuously | fast | ±12° |
| [12] | Semiconductor | 140 GHz | 42.8% | Discretely（7 beams） | fast | -45°~45° |
| [13] | Semiconductor | 1.95 THz | 20% | Continuously | fast | 40° |
| [14] | mechanical | 30 GHz | 22% | Continuously | slow | 60° |
| [15] | mechanical | 410 GHz | — | Continuously | slow | 30° |
| [16] | mechanical | 300 GHz | 30% | Continuously | slow | 86.2° |
| [17] | liquid crystal | 100 GHz | 8% | Continuously | fast | 55° |
| [18] | liquid crystal | 100 GHz | 8% | Continuously | fast | ±14° |
| [19] | Graphene | 1.095 THz | — | Continuously | fast | 13° |
| [20] | Graphene | 1.5THz | — | Continuously | fast | ±45° |
| This work | **MEMS mirrors** | **140 GHz** | **＞28.5%** | **Continuously** | **fast** | **30°/60°/90°** |

## 5. Conclusions

This article proposes an N-order terahertz MEMS beam scanning system with scalable beam scanning angle based on spatial beam focusing and multi-level tunable reflection array principles. Test and verify the first-order terahertz beam scanning system using a 7×8mm MEMS reflector and an 8×8mm focusing lens. Experiments have shown that this terahertz MEMS beam scanning scheme has the ability of large bandwidth and large scanning angle, while utilizing MEMS mirrors to achieve miniaturization and fast scanning. It has important application value in terahertz high-speed dynamic communication, high-precision scanning radar, high-resolution imaging, multi frequency domain imaging detection and other fields. We will continue to optimize and improve the scheme in the future, with a focus on improving the aperture utilization of the beam scanning system, continuously improving beam gain and beam shape.

**Author Contributions:** Conceptualization, W.Y. and H.P.; methodology, H.X.; investigation, Y.X. and H.L.; formal analysis, W.Y.; software, H.P.; data curation, H.P.; writing—original draft preparation, W.Y.; writing—review and editing, H.X.; supervision, H.L. M.L.and W.Y. All authors have read and agreed to the published version of the manuscript.

**Funding:**

**Data Availability Statement:** Not applicable.

**Conflicts of Interest:** The authors declare no conflict of interest.